\documentclass[preprint]{JASA}
\usepackage{amsmath,amssymb,amsfonts}
\usepackage{graphicx}
\usepackage{xcolor}
\usepackage{url}
\usepackage[T1]{fontenc}
\usepackage[utf8]{inputenc}
\usepackage[nolist]{acronym}
\usepackage{natbib}
\usepackage{siunitx}

\begin{document}

\title[]{EEG-Based Decoding of Sound Location: Comparing Free-Field to Headphone-Based Non-Individual HRTFs}

\author{Nils Marggraf-Turley}
\email{n.marggraf-turley23@imperial.ac.uk}
\affiliation{Dyson School of Engineering, Imperial College London, United Kingdom}

\author{Martha Shiell}
\email{mrhs@eriksholm.com}
\affiliation{Eriksholm Research Centre, Oticon A/S, Denmark}

\author{Niels Pontoppidan}
\email{npon@eriksholm.com}
\affiliation{Eriksholm Research Centre, Oticon A/S, Denmark}

\author{Drew Cappotto}
\email{d.cappotto@imperial.ac.uk}
\affiliation{Dyson School of Engineering, Imperial College London, United Kingdom}
 
\author{Lorenzo Picinali}			
\email{l.picinali@imperial.ac.uk}
\affiliation{Dyson School of Engineering, Imperial College London, United Kingdom}

\date{\today} 

\begin{abstract}
Sound source localization relies on spatial cues such as interaural time differences (ITD), interaural level differences (ILD), and monaural spectral cues. Individually measured Head-Related Transfer Functions (HRTFs) facilitate precise spatial hearing but are impractical to measure, necessitating non-individual HRTFs, which may compromise localization accuracy and externalization. To further investigate this phenomenon, the neurophysiological differences between free-field and non-individual HRTF listening are explored by decoding sound locations from EEG-derived Event-Related Potentials (ERPs). Twenty-two participants localized stimuli under both conditions with EEG responses recorded and logistic regression classifiers trained to distinguish sound source locations.

Lower cortical response amplitudes were observed for KEMAR compared to free-field, especially in front-central and occipital-parietal regions. ANOVA identified significant main effects of auralization condition ($F(1, 21) = 34.56$, $p < 0.0001$) and location ($F(3, 63) = 18.17$, $p < 0.0001$) on decoding accuracy (DA), which was higher in free-field and interaural-cue-dominated locations. DA negatively correlated with front-back confusion rates (r = -0.57, $p < 0.01$), linking neural DA to perceptual confusion.

These findings demonstrate that headphone-based non-individual HRTFs elicit lower amplitude cortical responses to static, azimuthally-varying locations than free-field conditions. The correlation between EEG-based DA and front-back confusion underscores neurophysiological markers' potential for assessing spatial auditory discrimination.  
\end{abstract}

\maketitle

\begin{acronym}
 \acro{USA}{United States of America}
 \acro{LTI}{Linear time-invariant}
 \acro{HRTF}{Head-Related Transfer Function}
 \acro{EEG}{Electroencephalography}
 \acro{ITD}{Interaural Time Difference}
 \acro{ILD}{Interaural Level Difference}
 \acro{ERP}{Event-Related Potential}
 \acroplural{ERP}[ERPs]{Event-Related Potentials}
 \acro{ML}{Machine Learning}
 \acro{KEMAR}{Knowles Electronics Mannequin for Acoustic Research}
 \acro{DA}{Decoding Accuracy}
 \acro{ICA}{Independent Component Analysis}
 \acro{EOG}{Electro-oculography}
 \acro{SVM}{Support-Vector Machine}
 \acro{MEG}{Magnetoencephalography}
 \acro{SNR}{Signal-to-Noise Ratio}
 \acro{SRM}{Spatial Release from Masking}
 \acro{HI}{Hearing Impaired}
 \acro{MMN}{Mismatch Negativity}
 \acro{VR}{Virtual Reality}
 \acro{SPL}{Sound Pressure Level}
 \acro{GFP}{Greater Field Power}
 \acro{FAL}{Fractional Area Latency}
 \acro{FF}{free-field}
 \acro{BRIR}{Binaural Room Impulse Response}
\end{acronym}
\acresetall 
\section{\label{sec:1} Introduction}

The human auditory system relies on spatial auditory cues extracted from \ac{ITD}, \ac{ILD}, as well as monaural spectral cues, to navigate acoustic environments. These cues not only enable the localization of sound sources but also facilitate more complex auditory tasks, such as segregating overlapping sounds in noisy settings \citep{cherryExperimentsRecognitionSpeech1953}. These spatial cues are highly individual, as they are shaped by the unique interaction of sound waves with each listener’s morphology. These individual spatial characteristics are encapsulated in the listener’s \ac{HRTF}, representing the spectro-temporal filtering of sounds from various locations around the head, capturing the physical interactions with their morphology as sound propagates to the ear canal. Consequently, \acp{HRTF} are used variously for creating virtual acoustic environments in \ac{VR} \citep{ganPersonalizedHRTFMeasurement2017},  consumer audio \citep{jotEfficientStructuresVirtual2017} and hearing aid technology \citep{eastgate3DGamesTuning2016}.  However, measuring \acp{HRTF} is time-consuming and requires specialised equipment, making it impractical for widespread application. Consequently, non-individual \acp{HRTF}, such as those derived from head-and-torso mannequins, are often employed which can lead to spatial inaccuracies due to the lack of individual spatial cues. Particularly common are front-back confusions where a source intended to be rendered in the front hemisphere is perceived as coming from the rear and vice versa \citep{wenzelLocalizationUsingNonindividualized1993}. This primarily occurs because non-individual HRTFs do not adequately simulate the spectral cues necessary for distinguishing locations where interaural cues are ambiguous \citep{wenzelLocalizationUsingNonindividualized1993}. Traditionally, listening tests are used to assess perceptual differences between \ac{HRTF} sets. However, such tests rely on the listener's ability to accurately communicate their perception, which often has low consistency \citep{kimInvestigationConsistencySubjective2020} and repeatability \citep{andreopoulou2016investigation}, and are not possible for individuals who cannot provide behavioral feedback (see also \cite{picinalilorenzoandkatzSystemtoUserUsertoSystemAdaptations2023}).

An alternative to perceptual tests is investigating spatial auditory perception from a neurophysiological perspective. \ac{EEG} is a non-invasive neuroimaging method that records the brain’s electrical activity by detecting voltage fluctuations on the scalp. One commonly analyzed EEG measure is the \ac{ERP}, a time-locked neural response reflecting sensory, cognitive, or motor processing associated with a specific stimulus or event. Consequently, \ac{EEG} can provide insights into the brain’s processing of auditory spatial cues without relying solely on subjective reports. By identifying neural correlates of the diminished perception associated with non-individual HRTFs, we can better understand the underlying mechanisms and potentially develop methods to enhance the effectiveness of these HRTFs. Furthermore, recent advancements in \ac{EEG} technology have allowed for its integration into devices that  
offer spatial audio processing. These include \ac{VR} headsets \citep{bernalGaleaPhysiologicalSensing2022, moinnereauInstrumentingVirtualReality2022},  prototypes of \ac{EEG} assisted hearing aids \citep{bechchristensenEEGAssistedHearingAids2018}, and patents for integrations into consumer earphones \citep{azemi2023biosignal}. This integration presents an opportunity for dynamic neurophysiological assessment of spatial auditory perception, potentially overcoming the limitations of traditional methods. As such, prior research has utilised EEG to assess various aspects of spatial auditory perception, including externalization \citep{colasSoundExternalizationDynamic2023}, immersion \citep{nicolEEGMeasurementBinaural2019} and presence \citep{zhangInvestigatingPotentialUse2022}.

Previous studies have also examined cortical responses to different forms of degraded virtual spatial cues, such as the use of non-individual HRTFs, and have observed significant differences in cortical activity between these conditions. \citet{palomakiSpatialProcessingHuman2005} investigated cortical response differences to localization with individual \acp{HRTF}, non-individual \acp{HRTF}, and combined or isolated \ac{ILD} and \ac{ITD} cues using \ac{MEG}. They reported that amplitude variations of the right-hemisphere N1m response were most pronounced for individual HRTFs and least for isolated \ac{ILD} cues and the systematicity of these right-hemisphere amplitude changes was correlated with localization accuracy. In the oddball paradigm of \citet{schrogerInterauralTimeLevel1996}, combined \ac{ILD} and \ac{ITD} cues elicited a greater \ac{MMN} response than isolated interaural cues. Using fMRI, \citet{callanNeuralCorrelatesSound2013} found that individual \acp{HRTF} elicited stronger activation in the bilateral posterior temporal gyri than internalised stereo stimuli. In the context of speech segregation, \citet{dengImpoverishedAuditoryCues2019a} found that individual \acp{HRTF} elicited stronger neural markers of selective auditory attention compared with impoverished cues (individual \ac{ILD}, or non-individual \ac{ILD}). Finally, \citet{wisniewskiEnhancedAuditorySpatial2016} investigated the neural correlates of \ac{HRTF} individualization using a roving oddball paradigm in which stimulus elevation changed. They found that elevation changes under individual \ac{HRTF} listening elicited a larger P3 response at posterior electrodes than non-individual \ac{HRTF} listening, and that this P3 amplitude difference was correlated with the proportion of correct behavioral elevation change detections.

Collectively, these studies highlight cortical response differences due to various degrees of spatial cue degradation, with degraded spectral cues typically eliciting delayed cortical responses with lower amplitude. However, several of the aforementioned studies use headphone-based individual \ac{HRTF} listening as their auralization condition with the least degraded spatial cues. Perceptual differences between listening with individual \acp{HRTF} and real acoustic conditions have been observed \citep{bronkhorstLocalizationRealVirtual1995, wightmanHeadphoneSimulationFreefield1989} and thus, the neural response variations reported in prior research may not fully represent variations to physical acoustic spatial hearing. While comparisons of cortical processing under real and virtual acoustic conditions have been conducted in animals \citep{behrendNeuralResponsesFree2004, kellerHeadrelatedTransferFunctions1998, campbellInterauralTimingCues2006}, such direct comparisons in humans are rare with only two studies reporting on this difference. \citet{getzmannEffectsNaturalArtificial2010} investigated the cortical response elicited by motion under free-field, non-individual \acp{HRTF}, and isolated \ac{ITD} or \ac{ILD} variations. Higher N1 and P2 amplitudes were found for free-field listening than for non-individual \ac{HRTF}. \citet{stodtComparingAuditoryDistance2024} compared cortical responses to distance changes between physical acoustic and non-individual \ac{HRTF} conditions using an odball paradigm, finding slight differences in P3a latency of the \ac{MMN} between physical and non-individual \ac{HRTF} stimuli. 

There remains a need to investigate cortical response differences directly with the ground-truth auralization condition of free-field listening. 
Whilst \citet{stodtComparingAuditoryDistance2024} and \citet{getzmannEffectsNaturalArtificial2010} have explored these conditions with sound source distance changes and motion, the effect on static sound source localization remains to be seen. Moreover, the correlations between cortical response variations and behavioral outcomes \citep{palomakiSpatialProcessingHuman2005, wisniewskiEnhancedAuditorySpatial2016, bialasEvokedResponsesLocalized2023} indicate a promising avenue for translating neural markers into quantitative measures of localization accuracy. This raises an important question: Can front-back confusions, common under non-individual HRTFs, be detected in neural responses? 

Sound source location decoding approaches show particular promise for this task of translating neurophysiological markers into a behavioral measure. These methods typically involve training a classifier to distinguish the spatial locations of sounds based on their corresponding \acp{ERP}. Previous work has shown that decoding is possible for distinguishing between locations on the horizontal \citep{bednarDifferentSpatiotemporalElectroencephalography2017} and median planes \citep{bialasEvokedResponsesLocalized2023} from free-field elicited \acp{ERP}. \cite{bednarDifferentSpatiotemporalElectroencephalography2017} trained \acp{SVM} on ERP data to decode sound source locations, finding that decoding accuracy reached significance later when distinguishing between spectral cue-dominated locations than interaural cue-dominated locations. In a similar paradigm, \citet{bialasEvokedResponsesLocalized2023}  used logistic regression models to decode median plane ERPs, reporting a correlation between decoding accuracy and behavioral localization accuracy. 

A preliminary study was conducted by \citet{marggraf-turleyDecodingSoundSource2024} using a multivariate decoding approach similar to that of \citet{bialasEvokedResponsesLocalized2023, bednarDifferentSpatiotemporalElectroencephalography2017}. This study served as a proof-of-concept, testing whether sound source location decoding was possible under non-free-field conditions. Initial indications suggested lower decoding accuracy for non-individual \ac{HRTF}-elicited \acp{ERP}, delayed significant decoding accuracy between free-field and non-individual \ac{HRTF} conditions, and a correlation between decoding accuracy and behavioral front-back confusion rates. However, the preliminary study involved only a single participant in each experiment, and the significance of these results was not established. 

The present work aims to investigate cortical response differences between localizing under free-field and non-individual HRTF auralization conditions and expand upon the preliminary decoding results of \citet{marggraf-turleyDecodingSoundSource2024} using a modified experimental paradigm and a larger cohort of participants. 

Considering the weaker cortical responses elicited by degraded spatial cues \citep{palomakiSpatialProcessingHuman2005, wisniewskiEnhancedAuditorySpatial2016, callanNeuralCorrelatesSound2013, dengImpoverishedAuditoryCues2019a} and the lower neural response amplitude to motion onset in virtual listening compared to free-field listening \citep{getzmannEffectsNaturalArtificial2010}, we propose the following hypothesis:
\begin{enumerate}
\item[$H_1$] Cortical responses elicited by a non-individual \ac{HRTF} during static sound source localization will have a lower amplitude than those elicited by free-field listening.
\end{enumerate}

Given the significant \ac{DA} observed in free-field location decoding studies \citep{bednarDifferentSpatiotemporalElectroencephalography2017, bialasEvokedResponsesLocalized2023}, alongside the increased front-back confusions for non-individual \ac{HRTF} listening \citep{wenzelLocalizationUsingNonindividualized1993} for locations where interaural cues are ambiguous, we hypothesise that:
\begin{enumerate}
\item[$H_2$] Sound source location \ac{DA} will be lower for non-individual \ac{HRTF} listening when decoding between spectral cue-dominated locations than for free-field.
\end{enumerate}

Based on previous findings that spatial cue degradation can result in increased latency of certain components of the \ac{ERP} \citep{palomakiSpatialProcessingHuman2005, stodtComparingAuditoryDistance2024} we further hypothesise that:
\begin{enumerate}
\item[$H_3$] Significant \ac{DA} will be reached later for non-individual \ac{HRTF} listening then for free-field.
\end{enumerate}

Finally, building on \citet{bialasEvokedResponsesLocalized2023}, who showed a correlation between decoding accuracy and free-field median plane localization accuracy, and the increased front-back confusions observed under non-individual \ac{HRTF} listening \citep{wenzelLocalizationUsingNonindividualized1993}, we propose that:
\begin{enumerate}
\item[$H_4$]  \ac{DA} between spectral-cue-dominated locations will correlate with behavioral front-back confusions.
\end{enumerate}

\section{\label{sec:2} Methods}
\subsection{\label{subsec:2:1} Apparatus, Stimuli and Procedure}
The experiment paradigm included a localization test using either free-field or non-individual \ac{HRTF} rendered stimuli, during which \ac{EEG} was measured. The participant cohort consisted of 22 subjects, divided into the following age groups: 18–27 years (n=11), 28–37 years (n=9), 48–57 years (n=1), and 58 years and above (n=1). Participants were seated on a height-adjustable chair at the centre of a loudspeaker array in a near-anechoic chamber. Each loudspeaker was positioned 1.4 meters from the participant. A custom chin-rest was used to adjust each participant’s head's height, aligning their ear canals with the center of the array. Participants were instructed to remain in the headrest to minimize head movements during the experiment.

Stimuli were delivered using an adapter-probe paradigm, similar to the methodology of \citet{bialasEvokedResponsesLocalized2023}, leveraging neural adaptation \citep{bendaNeuralAdaptation2021} to elucidate the brain’s response to changes in sound location and attenuate responses to general sound onset. 

\begin{figure}
    \centering
    \includegraphics[width=0.5\linewidth]{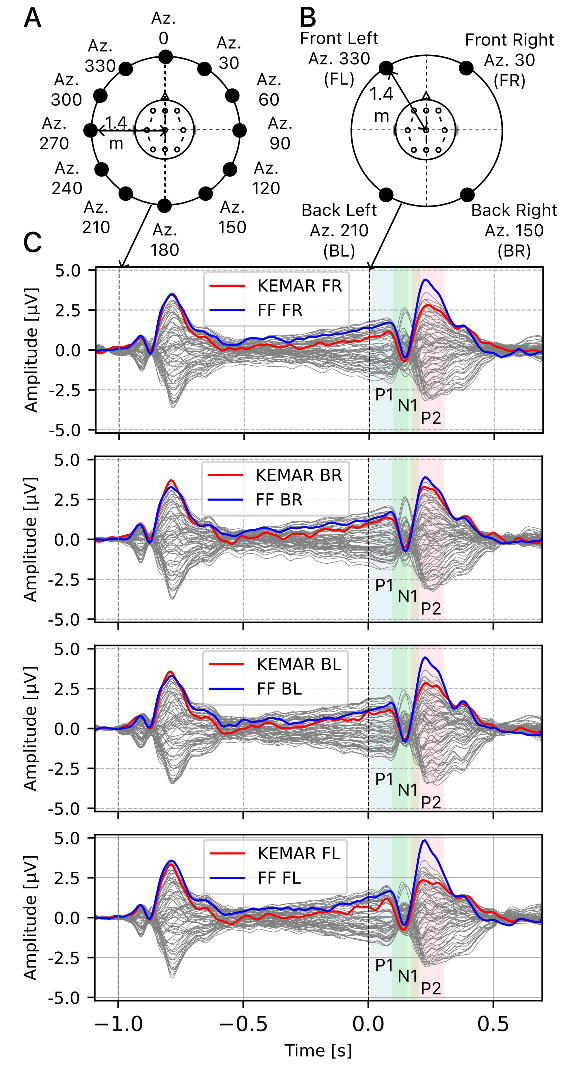}

    \caption{(color online) Experimental paradigm and \ac{ERP} plots. A and B show adapter and probe locations respectively. C shows average \acp{ERP} across both auralization conditions for each electrode in grey, with \acp{ERP} at electrode FCz highlighted for KEMAR (blue) and free-field (red) conditions. Each probe location is shown on a separate plot}
    \label{fig:method}
\end{figure}

In contrast to \citet{bialasEvokedResponsesLocalized2023}, where the adapter was delivered through two loudspeakers near the ears—leading to an internalised adapter and external probe—the adapter in this study was a 1000 ms pseudo-decorrelated white noise signal played from 12 locations, spaced 30 degrees apart on the horizontal plane (see Figure \ref{fig:method} A). This modification aimed to mitigate confounds related to changes in sound externalization. The probe was a 100 ms white noise burst delivered from one of four locations symmetrically arranged around the median and interaural planes, allowing approximately independent examination of interaural and spectral cues. The azimuth of these locations was 330$^\circ$, 210$^\circ$, 150$^\circ$, 30$^\circ$, referred to from here on as \textit{Front Left} (FL), \textit{Back Left} (BL), \textit{Back Right} (BR) and \textit{Front Right} (FR). A 1 ms equal-power crossfade was applied during the transition from adapter to probe to prevent transient artefacts. To reduce anticipatory cognitive processes related to sound onset timing, stimuli were presented with random offset-to-onset intervals ranging from 750 ms to 950 ms.

All stimuli were presented at 65 dB \ac{SPL}. Stimuli were generated using Python and output through a MOTU 24Ao audio interface, performing digital-to-analogue conversion at 48 kHz. The analogue signals were routed either to an array of custom-built single driver (Peerless PLS-P830987) loudspeakers or to Audiotechnica ATH-E50 insert earphones for binaural stimuli. Binaural stimuli were synthesised using the \ac{KEMAR} \ac{HRTF} from the SONICOM database \citep{engelSONICOMHRTFDataset2023}, recorded in the same room as this experiment. Head-tracking was omitted to avoid interference with \ac{EEG} electrodes. When specific loudspeaker locations did not match available \ac{HRTF} data, the 3DTI toolkit \cite{cuevas-rodriguez3DTunetoolkitOpensource2019} was used to interpolate the required spatial locations using barycentric interpolation among the nearest three head-related impulse responses. Equal loudness between free-field and binaural stimuli was ensured by adjusting the gain of both conditions to produce an equal \ac{SPL} from the listening position, measured using MiniDSP ears \citep{dspMiniDSPEARSUSB2017}.

To minimise \ac{EOG} artefacts, participants were instructed to focus their gaze on a point 1.4 m away at azimuth and elevation of 0$^\circ$. A total of 125 trials were recorded for each location and auralization condition. The experiment was divided into 20 blocks with KEMAR \ac{HRTF} stimuli and free-field conditions interleaved every 5 blocks. Participants were informed they could rest as long as they desired between blocks. The experiment lasted approximately 1 hour.

In 75\% of trials, a pure tone sounded 750 ms to 950 ms after the probe, prompting participants to indicate the perceived location of the previous probe using a number pad in front of them. Responses were restricted to four quadrants (Front Left, Front Right, Back Left, Back Right) and there was no response time limit. These behavioral localizations facilitated the investigation of correlations between neural and behavioral measures. 

\subsection{\ac{EEG} Recording and Preprocessing}

\ac{EEG} data were collected using a 64-channel BioSemi Active II system, adhering to the international 10-20 electrode placement system. The sampling rate was set at 2048 Hz, and the data were referenced to the average of all electrodes. Data preprocessing was performed using MNE-Python (version 1.2.0; \cite{gramfortMEGEEGData2013}) along with custom scripts.

To eliminate slow drifts and high-frequency noise while preserving relevant frequency bands, the EEG data were band-pass filtered between 0.1 Hz and 40 Hz following \citet{colasSoundExternalizationDynamic2023}. After filtering, the continuous EEG recordings were segmented into epochs ranging from -100 ms to 600 ms relative to the onset of the probe stimuli following \citet{bednarDifferentSpatiotemporalElectroencephalography2017}, based on triggers time-locked with each probe's onset. 

Following epoching, the EEG data were downsampled from 2048 Hz to 128 Hz. To manage artefacts, the \texttt{AutoReject} module \citep{jasAutorejectAutomatedArtifact2017} was utilised to automatically compute rejection thresholds specific to each channel. Any epochs that exceeded these thresholds were either repaired or excluded from further analysis. Subsequently, \ac{ICA} was applied to eliminate \ac{EOG} artefacts. The epoched data were decomposed into 30 components, and the components that correlated with \ac{EOG} activity were identified using electrode Fp1 as a reference. The identified artefact components were excluded by zeroing their contributions, and the EEG data were reconstructed from the remaining components. Since sound propagation to the ear canal took longer for the free-field condition, free-field epochs were shifted by $1.4/343 * 1000 = 4.1$ ms back in time.

\section{Results}
\subsection{ERP Analysis}

\begin{figure*}[htb]
    \centering
    \includegraphics[width=\textwidth]{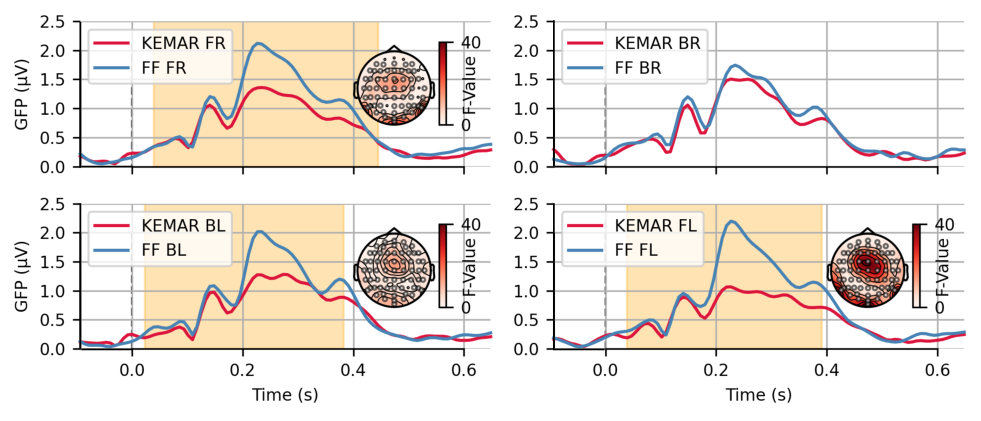}
    \caption{(color online) Cortical response comparisons between KEMAR and free-field (FF) conditions at four horizontal plane locations Az. 330$^\circ$ (FL), 210$^\circ$ (BL), 150$^\circ$ (BR), 30$^\circ$ (FR). \ac{GFP} plots show temporal dynamics between KEMAR (red) and FF (blue) conditions. Shaded areas mark the most significant time windows, identified using cluster-based permutation testing based on average F-scores. Topographical maps depict the spatial distribution of F-values averaged over these significant periods, with white dots indicating the sensors contributing to the clusters. Color bars represent the magnitude of the F-values, with darker regions highlighting areas of greatest statistical difference.}
    \label{fig:clusters}
\end{figure*}

\begin{figure*}[htb]
    \centering
    \includegraphics[width=\textwidth]{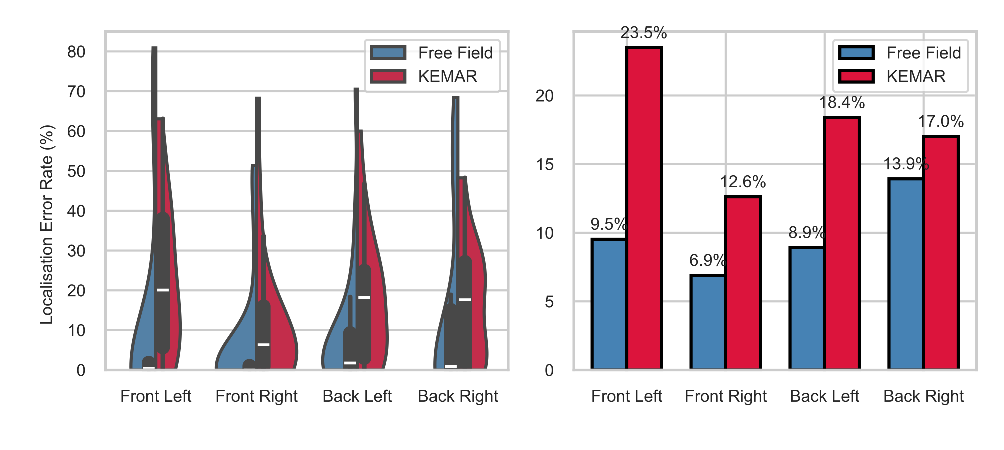}
    \caption{(color online) localization error rates across spatial locations for free-field and KEMAR conditions. A shows the distribution of localization error rates collapsed across all subjects for Front Left, Front Right, Back Left, and Back Right locations. Panel B presents the proportion of mislocalized locations in all \ac{EEG} epochs}
    \label{fig:misidentification_proportions}
\end{figure*}

To investigate the topography and magnitude of cortical response variations between auralization conditions for the same stimuli locations, thereby investigating $H_1$, a cluster permutation test was conducted on all subjects' epochs \citep{marisNonparametricStatisticalTesting2007}. This analysis was chosen as an initial, exploratory approach to assess cortical response variations across the entire ERP. The cluster permutation test was implemented with 10,000 permutations to detect significant \textit{(p<0.05) }spatiotemporal clusters, controlling for multiple comparisons by focusing on cluster-level statistics. Figure \ref{fig:clusters} displays the clusters with the highest average F-score between auralization conditions for each location, alongside the \ac{GFP} for the electrodes showing significant differences. The proportion of the epochs which comprised mislocalized perceptions is shown in Figure \ref{fig:misidentification_proportions}, along with the mislocalization rate across subjects.

Significant clusters were observed between auralization conditions for all locations except BR. Clusters began at 23-39 ms post-stimulus onset persisting until 383-445 ms. The \ac{GFP} was approximately ($\approx$\SIrange{0.5}{1}{\micro\volt} higher for the free-field condition than KEMAR for FL, FR and BL conditions, peaking at approximately 210 ms. The topographical F-maps in Figure \ref{fig:clusters} depict the distribution of F-values across the scalp, reflecting the variance between the free-field and KEMAR \ac{HRTF} conditions.  Darker regions in the F-map signify areas where the cortical responses differ more substantially between the two conditions, particularly in the front-central and occipital-parietal regions. No significant clusters were found for the BR condition, with \ac{GFP} for both KEMAR and free-field conditions peaking at (\SI{1.5}{\micro\volt} at approximately 210 ms.

Following this broad analysis, the effects of auralization condition and spatial location on the latency of specific components of the \ac{ERP} was conducted (P1, N1, and P2) at electrode FCz. FCz was chosen as it was the electrode with the highest average F-score in the significant clusters of Figure \ref{fig:clusters} for the majority of probe locations. Similar to the approach of \citet{stodtComparingAuditoryDistance2024}, the \ac{FAL} method was employed to investigate latency, as it provides a robust measure that is less susceptible to high-frequency noise and variability in peak amplitudes. The FAL method calculates the point at which the ERP waveform reaches 50\% of its total area within a predefined time window. Specific time windows were considered for each component: 10–150 ms for P1, 50–200 ms for N1, and 150–350 ms for P2.

The resulting FAL scores were subjected to repeated-measures ANOVAs with within-subjects factors of listening condition (KEMAR vs.\ free-field) and spatial location 
(\textit{FR, FL, BL, BR}). No significant main effects or interactions for any of the ERP components was found. For the P1 component, the main effect of the listening condition was not significant ($F(1,21) = 0.55$, $p = 0.468$), nor was the main effect of spatial location ($F(3,63) = 1.78$, $p = 0.159$), or their interaction ($F(3,63) = 1.22$, $p = 0.310$). Similarly, for the N1 component, neither listening condition ($F(1,21) = 1.00$, $p = 0.330$) nor spatial location ($F(3,63) = 0.45$, $p = 0.721$) showed significant effects, and their interaction was not significant ($F(3,63) = 2.44$, $p = 0.073$). For the P2 component, no significant effects were found for listening condition ($F(1,21) = 2.31$, $p = 0.144$), spatial location ($F(3,63) = 0.41$, $p = 0.748$), or their interaction ($F(3,63) = 0.87$, $p = 0.461$).

\subsection{Whole ERP Decoding}

To address $H_2$ and $H_4$, multivariate classification was implemented to classify \acp{ERP} into stimulus location following \citet{bednarDifferentSpatiotemporalElectroencephalography2017, bialasEvokedResponsesLocalized2023}. This decoding approach was used because it is sensitive to subtle variations in the neural response, as it considers the whole \ac{ERP} as opposed to individual components. 

Building upon the methodology of \citet{bialasEvokedResponsesLocalized2023}, logistic regression was implemented using \texttt{sklearn} \cite{scikit-learn}. The classifier was trained to distinguish between pairs of sound locations using \acp{ERP} recorded from 0 ms to 600 ms after probe onset. Data from all 64 electrodes were included during this interval, resulting in feature vectors of 64~$\times$77 dimensions per trial. 
To enhance the \ac{SNR}, every two epochs of the same condition were averaged. Datasets were balanced to ensure an equal number of data points for each participant (N=33) after averaging. The classification was conducted individually for each subject using 10-fold cross-validation.

To assess the statistical significance of \ac{DA}, a non-parametric permutation test was conducted following \citet{bednarDifferentSpatiotemporalElectroencephalography2017}. Class labels were randomly permuted, and the classifiers were retrained 1000 times to create a null distribution of \acp{DA}, with significance thresholds set by the extremes of this distribution.

For decoding, sound source locations were grouped into four pairs to approximately isolate interaural and spectral cue variations. The “Spectral Left” and “Spectral Right” pairs were defined as (FL, BL) and (FR, BR), respectively. Meanwhile, the “Interaural Front” and “Interaural Back” pairs were defined as (FL, FR) and (BL, BR), respectively. Table \ref{tab:significant_decoding_accuracy} presents the proportion of subjects achieving significant \ac{DA} for each location pair across auralization conditions. \ac{DA} was higher for interaural-cue-dominated location pairs for both auralization conditions than for spectral-cue-dominated locations. Notably, decoding between free-field frontal hemisphere interaural-cue-dominated locations (FF Interaural Front) was particularly successful, with 90.09\% of subjects achieving significant \ac{DA}. Conversely, spectral-cue-dominated decoding under the KEMAR condition was least successful, with only 9.09\% and 22.72\% of subjects reaching significance for the right and left hemispheres, respectively.

To examine the influence of auralization condition and location on decoding accuracy, a two-way repeated measures ANOVA was conducted with \textit{Auralization Condition} and \textit{Location pairs} as within-subject factors. A Shapiro-Wilk test on the residuals confirmed the assumption of normality ($W = 0.996$, $p = 0.94$). The ANOVA revealed significant main effects of \textit{Auralization Condition} ($F(1, 21) = 34.56$, $p < 0.0001$), and \textit{Location} ($F(3, 63) = 18.17$, $p < 0.0001$), indicating that both the auralization condition and the type of spatial cue significantly affect decoding accuracy. However, the interaction between \textit{Auralization Condition} and \textit{Location} was not significant, $F(3, 63) = 2.19$, $p = 0.1$.

\textit{Post hoc} paired t-tests with Holm-Bonferroni correction were conducted to examine the main effects of \textit{Auralization Condition}. Significant differences were observed for Spectral Left ($t = -2.73$, $p < 0.05$) and Spectral Right ($t = -2.81$, $p < 0.05$). Additionally, a significant difference was found between Interaural Front ($t = -6.11$, $p < 0.0001$). However, no significant difference was observed for Interaural Back ($t = -1.69$, $p > 0.05$). Regarding the main effect of \textit{Location}, significant differences were observed between KEMAR conditions: Spectral Left and Interaural Back ($t = -3.73$, $p < 0.05$), as well as between  Spectral Right and Interaural Front ($t = -3.67$, $p < 0.05$) and Spectral Right and Interaural Back ($t = -3.85$, $p < 0.01$). For the free-field conditions, significant differences were found between Spectral Left and Interaural Front ($t = -6.05$, $p < 0.0001$), Spectral Right and Interaural Front ($t = -7.11$, $p < 0.0001$).

\begin{figure*}[htb]
    \centering
    \includegraphics[width=\textwidth]{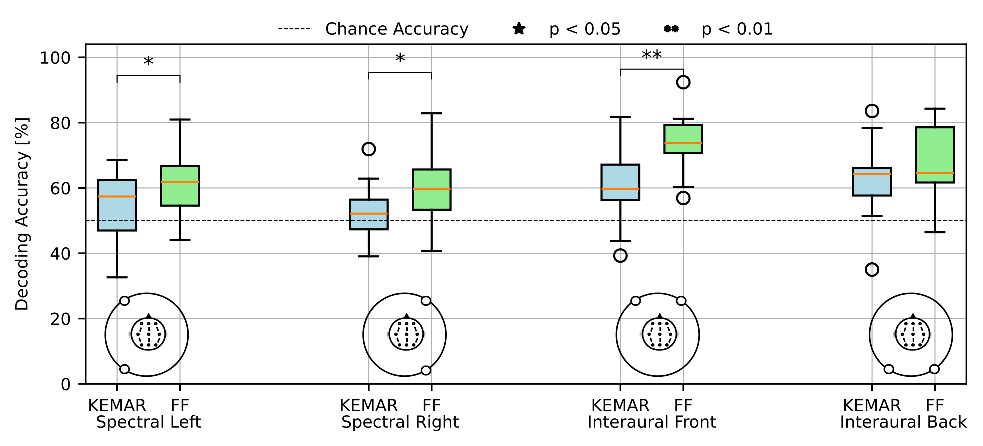}
    \caption{(color online) Box plots of decoding accuracy for pairwise sound source location comparisons based on \acp{ERP} elicited by KEMAR (blue) and free-field (FF) (green) stimuli. Decoding accuracy is presented for four location pairs, with the dashed line marking chance-level accuracy. Asterisks denote significance levels from post hoc tests on the main effect of auralization condition. Topographical insets below each pair show the relative positions of the compared locations.}

    \label{fig:boxplot}
\end{figure*}

Figure \ref{fig:boxplot} shows boxplot statistics for the distributions of \ac{DA} across all subjects, grouped by spatial location pair with significant interactions from post-hoc t-tests on the main effect of \textit{Auralization Condition} indicated.

\begin{table}[htb]
\centering
\caption{Percentage of subjects with significant decoding accuracy per location pair for free-field (FF) and KEMAR auralization conditions.}
\begin{tabular}{l c}
\hline
\textbf{Location Pair} & \textbf{No. Significant [\%]} \\
\hline
KEMAR Spectral L       & 22.72  \\
FF Spectral L          & 45.45 \\
KEMAR Spectral R       & 9.09  \\
FF Spectral R          & 27.27 \\
KEMAR Interaural Front & 36.36 \\
FF Interaural Front    & 90.90 \\
KEMAR Interaural Back  & 59.09 \\
FF Interaural Back     & 59.09 \\
\hline
\end{tabular}
\label{tab:significant_decoding_accuracy}
\end{table}

\subsection{Individual Time Point Decoding}

To evaluate \( H_2 \), a sliding-window decoding approach was employed, constructing separate classifiers at each time point in the \ac{ERP} across all electrodes. Similar to the whole-window decoding, a non-parametric permutation test was employed with 1000 permutations at each time point. This analysis aimed to capture the temporal dynamics of \ac{DA}, providing insights into the cortical processing delays associated with spectral and interaural-cue dominated location discrimination.

\begin{figure*}[htb]
    \centering
    \includegraphics[width=\textwidth]{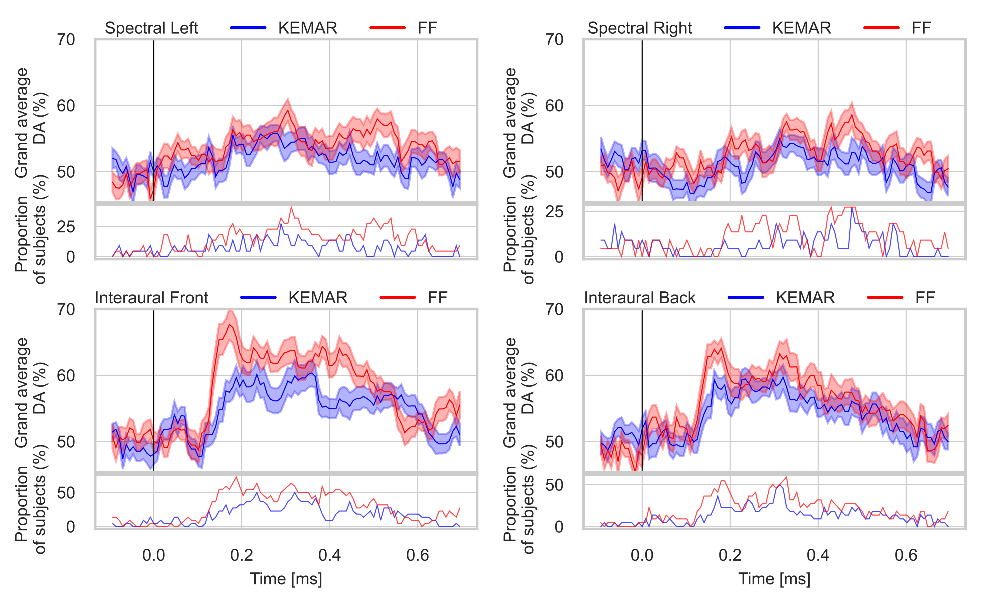}
    \caption{(color online) Temporal variation in pairwise decoding accuracy, averaged across subjects, for \ac{KEMAR} and free-field (FF) conditions. Shaded regions show $\pm SD$ estimated via bootstrapping. The lower subplots of each figure represent the proportion of subjects from whom significant \ac{DA} was reached at each time point obtained via permutation testing.}
    \label{fig:time_decoding}
\end{figure*}

Figure~\ref{fig:time_decoding} illustrates the temporal variation in pairwise decoding accuracy averaged across subjects, with each subplot comparing the performance between \ac{KEMAR} and free-field for a given location pair. The bootstrap method, following the approach of \citet{bialasEvokedResponsesLocalized2023}, was employed to resample decoding accuracies 10,000 times. This resampling provides an estimate of the group's average decoding accuracy and the uncertainty in this estimate, as indicated by the standard deviation of the resampled distribution. In addition to decoding accuracy, the proportion of subjects for whom significant \ac{DA} was achieved at each time point was calculated using the permutation tests. This proportion provides insight into the consistency of significant decoding across subjects over time.

For the interaural-cue-dominated location pairs, the bootstrapped decoding accuracy for both \ac{KEMAR} and free-field peaked at approximately 120 ms. For the interaural front locations, the peak proportion of subjects that reached significant \ac{DA} for both free-field and \ac{KEMAR} was >= 50\%. However, there was a slight latency in the peak proportion of subjects reaching significance for \ac{KEMAR} decoding, with its highest proportion occurring approximately 50 ms after that of free-field. Additionally, the proportion of subjects reaching significant decoding accuracy decayed faster for \ac{KEMAR} decoding than for free-field for the interaural front pair, decreasing sharply at approximately 400 ms from approximately 40\% to 10\%.

For the interaural back locations, two prominent peaks are observed in the proportion of subjects reaching significance, at approximately 190 ms and 300 ms, for both KEMAR and free-field decoding. 

In contrast, spectral-cue-dominated locations exhibited later peak DA and lower peak subject proportions compared with interaural-cue-dominated locations. For spectral left, the maximum proportion of significant subjects occurred at  $\approx300$ ms, and for spectral right, $\approx420$ ms. Here, peak subject proportions were lower, reaching approximately 30\% for both conditions. A latency in the proportion of significant subjects was evident for spectral right KEMAR decoding compared to free-field: the free-field condition reached $\approx20\%$ at 200 ms, while the first peak for KEMAR ($\approx18$\%) was not observed until 300 ms. However, DA peaks were less pronounced for KEMAR in spectral conditions, complicating latency interpretation.

\subsection{Decoding Accuracy and Front-back Confusion}
\begin{figure}[htb]
    \centering
    \includegraphics[width=\linewidth]{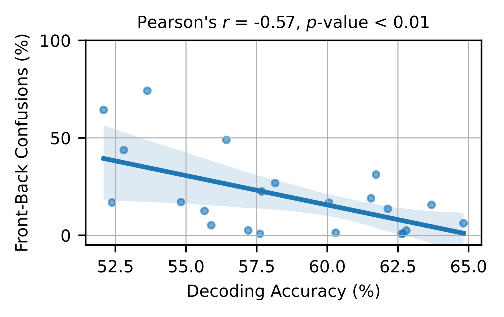}
    \caption{(color online) Front-back confusion rates and decoding accuracy for spectral-cue-dominated location pairs under KEMAR and free-field conditions. The regression line and shaded region, representing the 95\% confidence interval, show a significant negative correlation.}
    \label{fig:behavior}
\end{figure}

To address $H_4$, the relationship between behavioral front-back confusion rates and whole-window decoding accuracy was analyzed for spectral-cue-dominated location pairs. Figure \ref{fig:behavior} illustrates the correlation between front-back confusion rate and decoding accuracy for each subject averaged across spectral left and spectral right decoding, as well as both auralization conditions. The results reveal a moderate negative correlation (r = -0.57, p < 0.01).

\section{Discussion}

The primary objective of this investigation was to identify neurophysiological differences between localization under free-field and non-individual HRTF listening, and to establish a relationship between neural response and front-back confusion. To achieve this, an exploratory analysis was conducted to examine cortical response variations between the two auralization conditions, both topographically and temporally, using a spatiotemporal cluster permutation test. Upon identifying significant differences, an ANOVA was performed on \ac{ERP} component latencies. Subsequently, a location decoding paradigm was employed to translate these cortical response variations into behavioral outcomes. The findings from these analyses are discussed below, contextualized within the relevant literature, and potential improvements to the experimental paradigm and decoding methodology are proposed.

\subsection{Cortical Response Variations}

The initial spatiotemporal analysis revealed a significantly lower amplitude response in the EEG signal ($\approx$\SIrange{0.5}{1}{\micro\volt} GFP) for KEMAR stimuli compared to free-field stimuli for the FR, FL, and BL locations, thereby supporting hypothesis $H_1$. This attenuation in neural activity is consistent with the findings of \citet{getzmannEffectsNaturalArtificial2010}, who reported stronger cortical responses for free-field conditions relative to non-individual \ac{HRTF} at the onset of stimulus motion. Additionally, the absence of latency between free-field and non-individual \ac{HRTF} cortical response corroborates with \cite{getzmannEffectsNaturalArtificial2010}.

However, our results diverge from those of \citet{stodtComparingAuditoryDistance2024}, who observed comparable \ac{ERP} amplitudes for real and virtual conditions, alongside slight P3a latency variations. This discrepancy may be attributed to methodological differences between the studies. Specifically, \citet{stodtComparingAuditoryDistance2024} employed an oddball paradigm involving median plane distance changes, primarily altering spectral cues between stimuli. In contrast, our paradigm introduced differences in both interaural and spectral cues between the adapter and probe stimuli. Additionally, \citet{stodtComparingAuditoryDistance2024} conducted their experiments in a reverberant environment using \acp{BRIR} recorded with a HATS simulator, whereas our study was conducted in a near-anechoic room. This suggests that the presence of room reflections may mitigate neurophysiological differences between real and non-individual \acp{HRTF}, aligning with behavioral findings that indicate enhanced auditory perception in binaural conditions with reverberation compared to anechoic settings \citep{begault2001direct, angelDesignCanonicalSound2002}.

Notably, no significant differences were observed for the BR location between free-field and KEMAR \acp{ERP}. The BR location was the free-field position with the highest number of mislocalizations and exhibited the smallest discrepancy between free-field and KEMAR confusions. Additionally, the GFP for BR was approximately $0.5 \mu V$ lower than for other free-field locations. Previous studies have demonstrated that higher P3 amplitudes are associated with increased localization accuracies \citep{wisniewskiEnhancedAuditorySpatial2016}, and correlations exist between N1m amplitude variations and localization accuracy \citep{palomakiSpatialProcessingHuman2005}. As such, it seems that the confused BR stimuli may have resulted in weaker cortical encoding of spatial location, causing a lower amplitude neural response. The increased number of such confused epochs may then have led the spatiotemporal cluster permutation test to find no significant difference between auralization conditions.

In addition to spatiotemporal cluster permutation testing, the latencies of P1, N1, and P2 \ac{ERP} components were investigated using repeated-measures ANOVAs with within-subjects factors on \ac{FAL} values for each component. \citet{palomakiSpatialProcessingHuman2005} found that N1m latencies did not differ between individual and non-individual HRTFs. Our results show that the same holds between free-field and non-individual listening.
The absence of latency differences also supports analogous findings in the motion onset study by \citet{getzmannEffectsNaturalArtificial2010}, in which no latency differences were observed between free-field and non-individual \ac{HRTF} conditions. 

\subsection{Whole ERP Decoding}

Significant \ac{DA} was achieved for more than 50\% of participants for interaural-cue-dominated locations in both auralization conditions, except for the KEMAR interaural front location (36.36\%). \ac{DA} was also consistently higher for free-field conditions compared to KEMAR. This finding serves as a proof-of-concept that cortical responses, despite having a lower amplitude in the KEMAR condition, are still amenable to location decoding using non-individual \acp{HRTF}. Additionally, the $60^\circ$ separation between interaural-cue-dominated locations indicates a finer spatial resolution of decoding for free-field conditions compared to $180^\circ$ by \cite{bednarDifferentSpatiotemporalElectroencephalography2017}. However, decoding performance was diminished for spectral-cue-dominated location pairs, mirroring trends reported by \citet{bednarDifferentSpatiotemporalElectroencephalography2017}. 

For both KEMAR and free-field listening conditions, \ac{DA} was significantly lower when decoding spectral-cue-dominated locations compared to interaural-cue-dominated locations. This aligns with findings by \citet{bednarDifferentSpatiotemporalElectroencephalography2017}, who reported lower \ac{DA} for decoding between (Az $0^\circ$, El $-13.6^\circ$) and (Az $180^\circ$, El $-13.6^\circ$) compared to decoding between (Az $270^\circ$, El $-13.6^\circ$) and (Az $90^\circ$, El $-13.6^\circ$) in free-field conditions. These \ac{DA} variations suggest a weaker representation of spectral cues in the cortex as compared to binaural cues.

Supporting hypothesis $H_2$, significantly lower decoding accuracy was observed for KEMAR when decoding spectral-cue-dominated location pairs. Interestingly, decoding accuracy (\ac{DA}) was also reduced for interaural-cue-dominated location decoding in the KEMAR condition. This result is counterintuitive, as behavioral studies typically show that localization performance for positions with distinct interaural cues is less affected by non-individual \acp{HRTF} compared to positions where interaural cues are ambiguous \cite{wenzelLocalizationUsingNonindividualized1993}.

The findings of \citet{sleeAlignmentSoundLocalization2014} offer a potential explanation, demonstrating that neurons in the nucleus of the brachium of the inferior colliculus of marmoset monkeys are sensitive to the alignment of spatial cues and encode more information when these cues are coherent. Non-individual \acp{HRTF} may disrupt this alignment, potentially leading to impaired integration of spatial information and thereby reducing decoding accuracy.

\subsection{Individual Time Point Decoding}

In addition to whole \ac{ERP} decoding, individual time point decoding was conducted to investigate any temporal differences in \ac{DA} between both auralization conditions and location pairs. There were distinct latency differences between peak \ac{DA} between interaural and spectral-cue dominated locations. Since classifiers were trained on time points of the \ac{ERP}, and 130 ms aligns approximately with the N1 component observed in our ERPs, it seems that components from N1 onwards are pertinent for differentiating interaural-cue-dominated locations. In contrast, variations in P2 onwards appear to be more pertinent for spectral-cue-dominated location decoding. The increased latency for spectral-cue-dominated location decoding compared with interaural-cue-dominated decoding aligns with free-field results from \citet{bednarDifferentSpatiotemporalElectroencephalography2017}.

Furthermore, \citet{bialasEvokedResponsesLocalized2023} found \ac{DA} peaked at $\approx$400 ms for free-field median plane decoding, and \cite{wisniewskiEnhancedAuditorySpatial2016} found that the P3 response between 300–500 ms was more sensitive to elevation changes for individual versus non-individual \ac{HRTF}. Since their stimuli primarily varied in spectral cues—similar to the interaural plane-symmetric sources used in our study—our observed DA peak between 300 and 450 ms corresponds well with their reported time frames.

In the context of these previous studies, our findings provide support for the notion of segregated cortical processing pathways for spectral and interaural auditory cues \citep{fujikiHumanCorticalRepresentation2002, kaiserSimultaneousBilateralMismatch2000}.  However, similar to \citet{bednarDifferentSpatiotemporalElectroencephalography2017}, latency effects should be interpreted with caution for spectral-cue dominated locations due to their lower \ac{DA}.

Inter-auralization-condition latency variations were also evident. The peak number of participants for whom significant decoding was achieved occurred later for KEMAR than free-field in both spectral right and interaural front location pairs, showing partial support for $H_3$. This may indicate that, as well as a lower whole \ac{ERP} \ac{DA} for KEMAR stimuli than for free-field, there seems to be a latency at which these locations become discernible from cortical activity. Consistent with the whole-window \ac{ERP} decoding results, the \ac{DA} for the \ac{KEMAR} condition was lower than that for the free-field condition. 

\subsection{Behavioral Correlation}
A significant correlation between front-back confusion rates and \ac{DA} was observed, indicating a relationship between behavioral location discrimination and neurophysiological responses. Specifically, lower \ac{DA} between two locations was associated with higher rates of front-back confusion between those locations, supporting $H_4$. This finding is consistent with \citet{bialasEvokedResponsesLocalized2023}, who reported that \ac{DA} between free-field, median plane locations reflected subjects’ elevation localization accuracy. Our results extend these findings to conditions involving non-individual \acp{HRTF} and free-field listening for sources positioned on the cones of confusion on the horizontal plane. Notably, this relationship is primarily driven by the absence of high \ac{DA} in participants with high front-back confusion rates. Conversely, some participants exhibited low \ac{DA} despite having low front-back confusion rates, as depicted in the bottom left quadrant of Figure \ref{fig:behavior}.

\subsection{Limitations and Outlook}

Some limitations of this study should be acknowledged. Firstly, a static experimental setup was used where participants’ head movements were restricted using a custom chin rest. This design was intended to minimise muscle artefacts in the EEG data and ensure alignment with the horizontal, median, and interaural planes of the loudspeaker array. However, even minor head movements could potentially improve the resolution of front-back confusion by introducing additional interaural cues that help disambiguate spatial locations \citep{wightmanResolutionFrontBack1999}. If head movements had been more effectively controlled, it is possible that lower \ac{DA} would have been observed for spectral-cue-dominated free-field locations.

Secondly, as highlighted by \citet{rotaruWhatAreWe2024}, decoding models are sensitive to confounding factors such as eye-gaze patterns, which could lead to erroneous conclusions. In this context, if participants’ gaze behavior differed between auralization conditions, it could result in location decoding that is influenced by eye-gaze direction rather than true auditory perception. Nevertheless, we think this scenario is unlikely since participants were instructed to maintain their gaze on a fixed location for all trials. However, the possibility of small \ac{EOG}, muscle, or attentional artefacts differing between auralization conditions cannot be fully ruled out as a confounding factor. Alternatively, free-field and KEMAR stimuli may have activated different attentional mechanisms, potentially contaminating the cortical responses. This, however, would not have affected any of the within-auralization condition results.

Finally, an inherent confounding factor exists due to the simultaneous variation in both listening environment (free-field vs. headphones) and HRTF individualisation (individual vs. non-individual) across auralization conditions. Therefore, observed differences in cortical responses could be attributed to either or both factors. Future research should isolate these variables by comparing free-field listening and headphone-based listening with both individual and non-individual HRTFs.

This study underscores the neurophysiological differences between free-field and non-individual \ac{HRTF} listening, demonstrating the feasibility of decoding sound location from non-individual \ac{HRTF}, as well as the relationship between \ac{DA} and front-back confusion rates in both spatial environments. One question that arises from our study is whether the observed correlation between \ac{DA} and front-back confusion rates extends to general localization accuracy across multiple sound sources. Moreover, it would be intriguing to examine the \ac{DA}-behavior relationship in hearing-impaired individuals, as their cortical responses to spatial cue processing are known to differ significantly \citep{hanEarSpecificHemisphericAsymmetry2021}. Lastly, exploring whether a classifier trained on \acp{ERP} data from a general population of listeners could successfully decode sound locations for an unseen participant is an interesting avenue. Such an advancement could negate the need for participant-specific classifiers, enabling the assessment of localization accuracy without preliminary calibration.

\section{Conclusion}

This study investigated the neurophysiological differences between free-field and headphone-based non-individual \ac{HRTF} (from KEMAR) sound source localization using EEG-based decoding. KEMAR stimuli elicited significantly lower amplitude cortical responses compared to free-field stimuli across most spatial locations. Significant \ac{DA} was achieved for most subjects for KEMAR decoding for interaural-cue-dominated locations, with lower \ac{DA} for spectral cue-dominated locations. A lower decoding accuracy was observed for all location pairs for KEMAR decoding than free-field reflecting the impairment non-individual \acp{HRTF} cause to the neural encoding of location.

Temporal analysis revealed qualitative latency shifts in cortical processing when using non-individual \acp{HRTF} for the frontal hemisphere interaural-cue-dominated location pair. Furthermore, a significant negative correlation between decoding accuracy and behavioral front-back confusion rate was found, highlighting the potential of \ac{ERP}-based decoding as a neurophysiological marker of front-back confusion.

Despite limitations such as restricted head movements and potential artefact confounds, this research enhances understanding of how non-individual \acp{HRTF} affect the neural mechanisms underlying sound localization. Future studies should incorporate individual \acp{HRTF}, explore dynamic listening conditions, and include more diverse participant groups to elucidate further the relationship between neurophysiological measures and behavioral spatial discrimination.

\section{Author Declarations}
\subsection{Conflict of Interest}
The authors have no conflicts to disclose.
\subsection{Ethics Approval}
This study was approved by the Research Governance and Integrity Team (RGIT) at Imperial College London. SETREC No. 6878818. Informed consent was obtained from all participants.
\section{Data Availability}
The data that support the findings of this study are available on request from
the corresponding author.
\bibliography{final_bib}

\end{document}